\documentclass{article}
\begin{document}
\newcommand{\be}{\begin{equation}}
\newcommand{\ee}{\end{equation}}
\newcommand{\ba}{\begin{eqnarray}}
\newcommand{\ea}{\end{eqnarray}}
\newcommand{\no}{\nonumber\\}
\newcommand{\vf}{\varphi}
\title{Special solutions for Ricci flow equation in 2D  using the 
linearization approach}
\author{Stefan Adrian Carstea
\thanks{E-mail:~~~ acarst@theory.nipne.ro}
\and
Mihai Visinescu 
\thanks{E-mail:~~~ mvisin@theory.nipne.ro}\\
{\small \it Department of Theoretical Physics,}\\
{\small \it National Institute of Physics and Nuclear Engineering,}\\
{\small \it Magurele, P.O.Box MG-6, RO-077125 Bucharest, Romania}}
\date{ }
\maketitle

\begin{abstract}
The $2D$ Ricci flow equation in the conformal gauge is studied using 
the linearization approach. Using a non-linear substitution of 
logarithmic type, the emergent quadratic equation is split in 
various ways. New special solutions involving arbitrary functions are 
presented. Some special reductions are also discussed.

~

Pacs: 02.30.Jk Integrable systems

~~~~~~~ 04.60.Kz Lower dimensional models

\end{abstract}

\section*{}
In the last time the Ricci flow equations become a major tool for 
addressing a variety of problems in physics \cite{B1,B2,B3} and mathematics 
\cite{CCC}. The Ricci flows are second order non-linear parabolic 
differential equations for the components of the metric $g_{\mu\nu}$ of 
an $n$-dimensional Riemannian manifold which are driven by the Ricci 
curvature tensor $R_{\mu\nu}$:
\be\label{R}
\frac{\partial}{\partial t} g_{\mu\nu} =  - R_{\mu\nu}\,.
\ee

This equation describes geometric deformations of the metric $g_{\mu\nu}$ 
with parameter $t$. In particular the Ricci flow equations on two 
dimensional manifolds have attracted considerable attention on the 
physical literature in connection with two-dimensional black hole 
geometry, exact solutions of the renormalization group equations that 
describe the decay of singularities in non-compact spaces, etc. 
\cite{B1}. 

As an attempt to quantize gravity, it is interesting to investigate 
quantum field theory in a curved space-time background. Solving 
relativistic field equations in $(3+1)$-dimensional curved space-time 
is generally a difficult process. An alternative approach is to 
consider lower-dimensional space-times models where exact solutions may 
be obtained. It is now long time since lower-dimensional gravity proved 
to exhibit many of the qualitative features of $(3+1)$-dimensional 
general relativity, high dimensional black holes, cosmological models 
and branes \cite{GKV,B}.

The purpose of this paper is to present new explicit solutions for $2D$ 
Ricci flow equation using a linearization approach.

In two dimensions it is useful to consider a local system of 
conformally flat coordinates in which the metric has the form
$$ d s^2_t = \frac{1}{2} e^{\Phi(x ,y; t)} (d x^2+ d y^2) = 
 2 e^{\Phi(z_+,z_-;t)} d z_+ d z_- $$
using Cartesian coordinates $x, y$ or the complex conjugate variables 
$2 z_\pm = y \pm i x$.

Having in mind that the only non-vanishing component of the Ricci 
tensor is
$$
R_{+-} = - \partial_{+} \partial_{-} \Phi(z_+,z_-;t) 
$$
the Ricci flow equation (\ref{R}) becomes
\be\label{ecphi}
\frac{\partial}{\partial t} e^{\Phi(z_+,z_-;t)}  =  
\partial_{+} \partial_{-} \Phi(z_+,z_-;t)\,.
\ee

In what follows we shall use the substitution
$$
v(z_+, z_-; t) = e^{\Phi(z_+, z_-; t) } 
$$
writing eq. (\ref{ecphi}) in the form
\be\label{2}
(v)_t = (\ln v)_{z_+ z_-} \,.
\ee

This equation has been studied in detail from the algebraic point of 
view in \cite{B3}. It is considered as a "continual" version of the 
general Toda-type equation for a given Lie algebra. Also in \cite{B3}
it is presented a formal power series solution by expanding the
path-ordered exponentials. Although the proposed general solution 
provides a formal complete solution for (\ref{2}), its form is quite 
intricate and difficult to handle.

Our approach to the equation is somewhat different. We use a direct 
nonlinear substitution, then split the resulting nonlinear equation. 
This results in a class of special solutions. Moreover we consider that 
the equation (\ref{2}) is rather in the class of {\it linearizable} 
systems than Lax-pair solvable ones. 

Our supposition comes from the following observations:
\begin{enumerate}
\item Starting with (\ref{2}) and using the substitution $v = 
\vf_{z_+}$, after integrating once with respect to $z_+$ we get
\be\label{2'}
\vf_t \vf_{z_+} = \vf_{z_+ z_-} + C \vf_{z_+}
\ee
where $C$ should be a function of $z_{-}$ and $t$, but for the 
moment it is considered constant. Now making the 
substitution $\vf = \ln F $ we will end up with the following quadratic 
equation:
\be\label{3}
F_t F_{z_+} - F_{z_+ z_-} F + F_{z_+} F_{z_-} - C F F_{z_+} = 0 \,.
\ee

This equation has the following multi-shock-like solution:
\be\label{4}
F = 1 + e^{\eta_1} + e^{\eta_2} + \cdots + e^{\eta_N}   
\ee
where $\eta_i = k_i z_+ - C(z_- - t)$ \,, for any $k_i$ and positive 
integer $N$. 
Of course the solution (\ref{4}) is {\it not} a general solution 
because (\ref{3}) is {\it not} a bilinear Hirota form (it is not 
gauge-invariant \cite{Ramani}) and $C$ is just a constant. But the 
lacking of interaction between exponentials is characteristic to 
linearizable systems (like Burgers, Liouville, etc.) \cite{CRG}.
\item Equation (\ref{2}) does {\it not} pass the Painlev\' e test. 
Usually Painlev\' e test is considered an integrability detector. Of 
course it is not infallible. There are many equations which do not pass 
the Painlev\' e test but still are completely integrable. These are 
either in the class of Hamiltonian systems with separable 
Hamilton-Jacobi equation, or are again some {\it linearizable} systems 
\cite{Trembley}.
\end{enumerate}

Accordingly we are going to seek an underlying linear (or solvable) 
system for (\ref{2}). Let us remark that eq. (\ref{2}) is symmetric in 
variables $z_+$ and $z_-$ and, consequently, in the rest of the paper, 
the role of these variables can be interchanged.

Let us assume that $C$ is no longer a constant, but a free function of 
$z_-$ and $t$. In this case, defining
$$
\vf = \psi + \int_{-\infty}^{t} C(z_-, t ) dt 
$$
then (\ref{2'}) will have the form
\be\label{5}
\psi_{z_+} \psi_t = \psi_{z_+ z_-}\,.
\ee
Using the same nonlinear substitution $\psi = \ln F$ we get
\be\label{6}
F_{z_+} (F_t + F_{z_-})  = F F_{z_+ z_-}\,.
\ee

Equation (\ref{6}) can be split in some linear or nonlinear solvable 
equations. Of course, all the possibilities we are going to analyze 
will give only special solutions and not general ones.

First of all, we shall split eq. (\ref{6}) into a system of linear 
equations. Here we list the possibilities:
\begin{itemize}
\item {\it Linearization Ia}
\ba
F_{z_+ z_-} &=& 0\no
F_t + F_{z_-} &=& 0 \nonumber
\ea
with the general solution
$$
F(z_+, z_-, ; t) = f(z_+) + g(t- z_-)
$$
where $f, g$  are {\it arbitrary functions}. The solution of (\ref{2}) 
is
\be\label{7}
v(z_+, z_-; t) = \frac{f'(z_+)}{f(z_+) + g(t-z_-)}
\ee
which is in fact the generalization of the multi-shock solution 
(\ref{4}).
\item {\it Linearization Ib}
\ba
F_{z_+} &=& F_{z_+ z_-}\no
F_t + F_{z_-} &=& F\,.\nonumber
\ea
This system is equivalent with the previous one by means of the 
transformation:
\be\label{8}
F(z_+, z_-; t) \longrightarrow e^{z_-} F(z_+, z_-; t)
\ee
and $v(z_+, z_-; t)$ is invariant.
\item {\it Linearization Ic}
\ba\label{III}
F_t + F_{z_+} &=& F_{z_+ z_-} \no
F_{z_+} &=& F\nonumber
\ea
with the general solution
$$
F(z_+, z_-, ; t)  = h(t-z_-) e^{z_+ + z_-} 
$$
for any arbitrary function $h$. Unfortunately this gives a trivial solution 
for (\ref{2}), namely $v = 1$.
\end{itemize}

The next attempt is to split eq. (\ref{6}) in a solvable system of 
nonlinear equations. Like in the linearization of the type I, we have 
the possibilities:

\begin{itemize}
\item {\it Linearization IIa}
\ba\label{IV}
F_{z_+} &=& F^\alpha \,,\forall \alpha \,,\no
F^{\alpha - 1}(F_t + F_{z_-}) &=& F \,.
\ea
The advantage of this splitting is that the first equation of the 
system (\ref{IV}) is a Bernoulli one with the solution
$$
F(z_+, z_-; t) = \{(1- \alpha)[z_+ + h(z_-,t)]\}^{\frac{1}{1-\alpha}} 
$$
where $h (z_-, t)$ is an arbitrary function. Introducing this 
expression in the second equation of the system one finds a linear 
equation for $h (z_-, t)$:
$$
h_t + (1- \alpha) h_{z_-} = 0\,.
$$

In this way, the general solution for the system (\ref{IV}) is
$$
F(z_+, z_-; t)=  \{ (1- \alpha) [ z_+ + C z_+ (t - \frac{z_-}{1-\alpha} 
)] \}^{\frac{1}{1-\alpha}}
$$
which gives
\be\label{9}
v(z_+, z_-,t) = \frac { 1 + C (t - \frac{z_-}{1-\alpha} )}
{z_+ + C z_+ (t - \frac{z_-}{1-\alpha}  )}  = \frac {1}{z_+}\,.
\ee
Accordingly this nonlinear splitting gives a stationary solution, i.e. 
independent of the parameter of deformation $t$.
\item {\it Linearization IIb}
\ba\label{V}
F_{z_+ z_-} &=& F^\alpha F_{z_+}\no
F_t + F_{z_-} &=& F^{\alpha + 1} \,.
\ea
From the first equation of the system we get
$$
F_{z_-} = \frac{1}{\alpha +1}F^{\alpha +1} + \beta(z_-, t)
$$
with $\beta$ an arbitrary function. Introducing in the second equation 
we get
$$
F_t + F_{z_-} = (\alpha + 1)(F_{z_-} - \beta)
$$
having the solution
$$
F(z_+, z_-; t) = - \frac {\alpha + 1}{\alpha} \int^{z_-} \beta (\xi, 
z_- + \alpha t - \xi) d\xi + h(z_+, z_- + \alpha t)\,.
$$

Now solving the first equation for $h$ in the case of a general 
function $\beta$ is a difficult task. In any case, for $\beta = 0$ the 
system can be solved having the solution
\be\label{10}
v(z_+, z_-; t) = \frac{f'(z_+)}{f(z_+) - \alpha (1 + \alpha) z_- + 
\alpha t}\,.
\ee
\end{itemize}

We remark that the nonlinear splitting of the bilinear form gives 
solutions which are less general than (\ref{7}).

A different approach to eq. (\ref{6}) can be done  using a special 
combination of the variables $z_+$ and $z_-$ as a new independent 
variable. For example the well known solutions of cigar-type, or rational 
type, discussed in \cite{B3} can be obtained by introducing the following 
substitution:
\be\label{11}
z_+ + z_- = \xi\,.
\ee
Then (\ref{2}) becomes
$$
v_t = (\ln v )_{\xi\xi} \,.
$$

This equation has been extensively studied by Rosenau in \cite{Rosenau}
where, using the "addition property",  he found cigar-type and rational 
type solutions.

Of course one can consider other symmetric combinations of $z_+$ and 
$z_-$ to give new independent variables. The general method would imply 
the Lie-point symmetries (to find all the similarity reductions) but we 
are not going to do this here. Rather we will consider the following 
substitution:

\be\label{12}
z_+ z_- = \xi \,.
\ee

Using the same machinery we end up with the following bilinear 
equation:
\be\label{13}
F_\xi (F_t + \xi F_\xi) = \xi F F_{\xi\xi}
\ee
and, of course 
$$v(z_+, z_-; t) = \frac{\partial}{\partial \xi} \ln F$$.

As in the linearization I, the following alternatives are obvious: 

\begin{itemize}
\item {\it Linearization IIIa}

Now, we can split (\ref{13}) in the same way as before:
\ba
&& F_t + \xi F_\xi =0\no
&& F_{\xi\xi} = 0 \nonumber
\ea
with the general solution 
$$F(\xi, t) = a \xi e^{-t} + b$$ 
where $a,b$ are constants. In this case
\be\label{14}
v (z_+, z_-; t) =  \frac{1}{z_+ z_- + \frac{b}{a} e^t}\,.
\ee

\item {\it Linearization IIIb}
\ba
&& F_\xi = \xi F_{\xi\xi}  \no
&& F_t + \xi F_\xi =F\nonumber \,.
\ea

From the first equation of this system we have 
$$F(\xi, t) = \frac{1}{2} \xi^2 h(t) + \mu(t)$$ 
and from the second $h(t) = a e^{-t} \,, \mu(t) = c e^t$ 
with $a,c$ constants. Accordingly
\be\label{15}
v (z_+, z_-; t)= \frac{z_+ z_-}{\frac{1}{2}(z_+ z_-)^2  + \frac{c}{a} 
e^{2t}}\,.
\ee
\end{itemize}

The next  three splittings give trivial or stationary solutions:

\begin{itemize}
\item {\it Linearization IIIc}
\ba
&& F_\xi = F_{\xi\xi}\no
&& F_t + \xi F_\xi = \xi F \,.\nonumber
\ea
We have $F = a e^\xi $ which leads to 
\be\label{16}
v(z_+, z_-; t) = 1\,.
\ee

\item {\it Linearization IIId}
\ba
&& F_\xi = \xi F \no
&& F_t + \xi F_\xi = F_{\xi\xi}  \nonumber
\ea
with $F (\xi , t) = a e^{\frac{1}{2} \xi^2 + t} $ and  we get
\be\label{17}
v(z_+, z_-; t) = z_+ z_- \,.
\ee
\end{itemize}

Finally, let us consider the following splittings of eq. (\ref{13}) in  
solvable systems of nonlinear equations: 

\begin{itemize}
\item {\it Linearization IVa}
\ba
&& F_\xi = F^\alpha\no
&& F^{\alpha -1} (F_t + \xi F_\xi)  = \xi F_{\xi\xi} \,.\nonumber
\ea
From the first equation of the system we get 
$$F = [ (1 - \alpha) (\xi +  g(t))]^{\frac{1}{1-\alpha}}\,,$$ 
but if we introduce it in the second 
equation we  have $g' + 1 = \alpha z$ which gives $g' = -1$ and $\alpha 
= 0$ and consequently $F$ is trivial.
\item {\it Linearization IVb}

Another incompatible splitting would be ($\alpha$ and $\beta$ are 
constants):
\ba
&& F_\xi = \xi^\beta F^\alpha\no
&& \xi^{\beta - 1} F^{\alpha -1} (F_t + \xi F_\xi)  = F_{\xi\xi} \nonumber
\ea
which has no solution.
\item {\it Linearization IVc}

Another choice could be
\ba
&& F_{\xi\xi} = F F_\xi\no
&& F_t + \xi F_\xi  = \xi F^2 \,.\nonumber
\ea
From the first equation of the above system we have 
$$F = h(t) \tan [\frac{1}{2} (\xi h(t) + g(t))]$$ 
with $h(t) $ and $g(t)$ arbitrary 
functions of $t$. From the second equation we obtain differential 
equations for $h$ and $g$ which are not compatible. Therefore we haven't
any solution in this case. 
\end{itemize}

To conclude, the linearization approach represents an efficient 
procedure to generate solutions of the Ricci flow equation (\ref{2}).
We remark that the solution (\ref{7}) is the most general and 
represents the largest linearizable sector.

The exhaustive study of the Lie symmetries and similarity reductions 
performed on the quadratic equation (\ref{6}) will be the subject of 
forthcoming work \cite{CV}.
\subsection*{Acknowledgments}
The authors have been partially supported by the MEC-AEROSPATIAL Program, 
Romania.

\bibliographystyle{amsplain}

\end{document}